\documentclass[reprint,amsmath,amssymb,aps,twocolumn,floatfix,superscriptaddress]{revtex4-1}

\usepackage{graphicx}
\usepackage{hyperref}
\usepackage{siunitx}
\usepackage{amsmath}

\newcommand{\ie}{\emph{i.\,e.}}

\begin{document}

\title{Nonequilibrium thermodynamics of erasure with superconducting flux logic}

\author{Olli-Pentti Saira}
\thanks{Corresponding authors: osaira@bnl.gov (present address), roukes@caltech.edu}
\affiliation{Condensed Matter Physics and Kavli Nanoscience Institute, California Institute of Technology, Pasadena, CA 91125}
\affiliation{Computational Science Initiative, Brookhaven National Laboratory, Upton, NY 11973}
\author{Matthew H. Matheny}
\author{Raj Katti}
\author{Warren Fon}
\affiliation{Condensed Matter Physics and Kavli Nanoscience Institute, California Institute of Technology, Pasadena, CA 91125}
\author{Gregory Wimsatt}
\author{James P. Crutchfield}
\affiliation{Complexity Sciences Center and Physics Department,
University of California at Davis, One Shields Avenue, Davis, CA 95616}
\author{Siyuan Han}
\affiliation{Department of Physics and Astronomy, University of Kansas, Lawrence, KS 66045}
\author{Michael L. Roukes}
\thanks{Corresponding authors: osaira@bnl.gov (present address), roukes@caltech.edu}
\affiliation{Condensed Matter Physics and Kavli Nanoscience Institute, California Institute of Technology, Pasadena, CA 91125}
\date{\today}

\begin{abstract}
We implement a thermal-fluctuation driven logical bit reset on a superconducting flux logic cell. We show that the logical state of the system can be continuously monitored with only a small perturbation to the thermally activated dynamics at 500~mK. We use the trajectory information to derive a single-shot estimate of the work performed on the system per logical cycle. We acquire a sample of $10^5$ erasure trajectories per protocol, and show that the work histograms agree with both microscopic theory and global fluctuation theorems. The results demonstrate how to design and diagnose complex, high-speed, and thermodynamically efficient computing using superconducting technology.
\end{abstract}

\maketitle

Information storage and processing are vital in coordinating modern society. A considerable fraction (10\%) of the global electrical power output is spent on operating and cooling the required computing infrastructure~\cite{Mills2013}. On scales large and small, reduction and mitigation of the processor waste heat is critically important to high-performance computing. Two complementary strategies for developing an optimal computing platform~\cite{Lloyd2000} suggest themselves. The first improves the speed and energy-efficiency of the hardware platforms through engineering advances, and the second, a scientific endeavor, identifies and pursues the fundamental physical limits of computing machines. The latter originates most directly in the works of Landauer~\cite{Landauer1961}, who argued from a microscopic perspective that logically irreversible operations have an irreducible energy cost. This limit is approached, though, only when the clock rate of the computation is low enough to allow nearly-adiabatic physical evolution~\cite{Berut2012,Jun2014,Hong2016,Gaudenzi2018,Yan2018,Hofmann2016}. Most generally, physically-embedded computing requires a trade-off between between efficiency and speed, amongst other factors~\cite{Gopalkrishnan2016}.

\begin{figure}[t!]
    \includegraphics[width=3.375in]{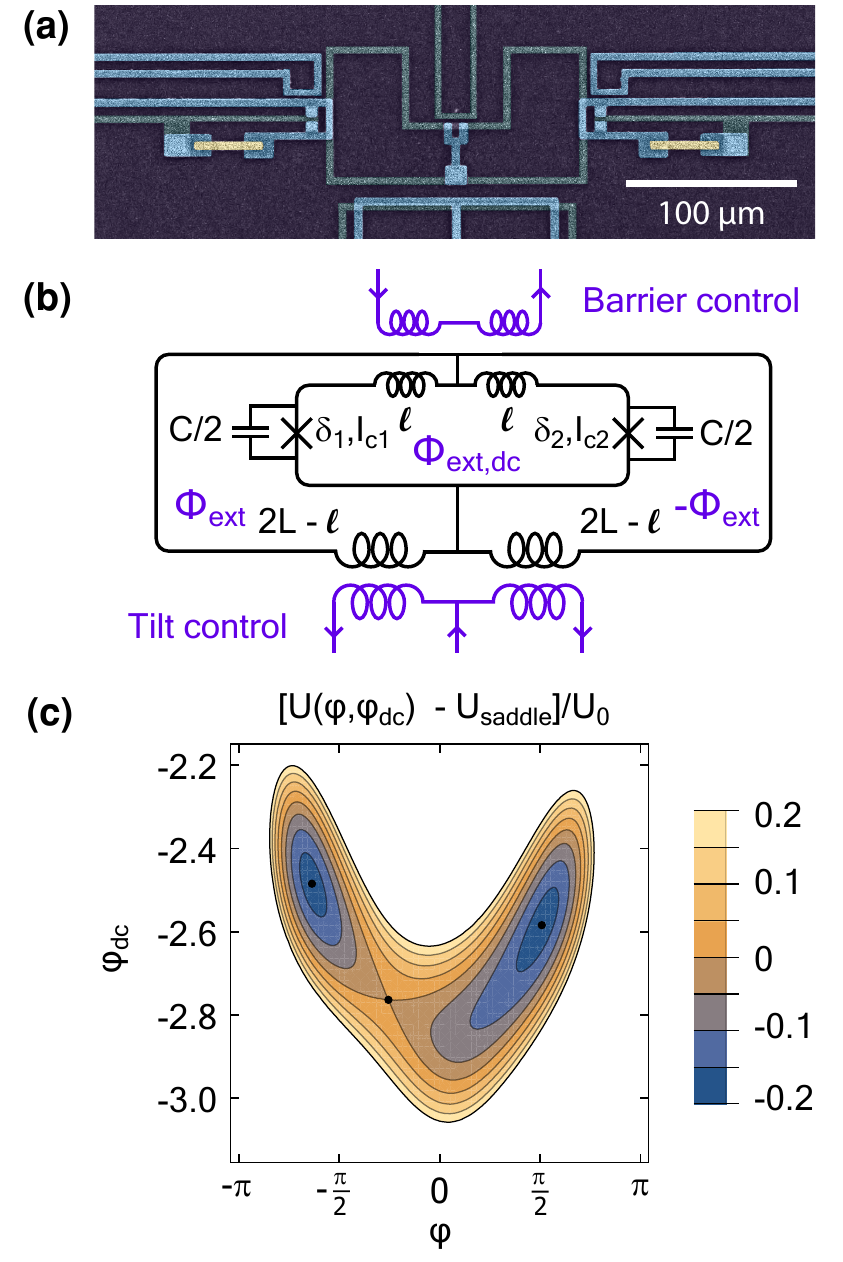}
    \caption{Gradiometric flux logic cell. (a) False-color electron micrograph of the device, realized as a two-layer superconducting circuit on an insulating silicon substrate. (b) Simplified circuit schematic of the information-bearing subsystem. We take the dynamical coordinates to be the total magnetic fluxes $\varphi$ and $\varphi_{dc}$ threading the loops. (c) Contour plot of the potential calculated for the component values of the studied device, and external bias fluxes $(\varphi_x, \varphi_{x,dc}) = (-0.1018, -2.5887)$ coinciding with the start of the bit erasure protocols studied later. Two local, metastable minima, and the unique saddle point are marked (black dots).}
    \label{fig:1}
\end{figure}

A key advance to efficient nonadiabatic computing appeared with the fluctuation theorems (FTs) that exactly describe the thermodynamics of small systems -- systems that are necessarily driven out of equilibrium by external controls during information processing~\cite{Jarzynski1997,Crooks1999}. Experimental tests of FTs have been performed in a variety of microscopic systems~\cite{Liphardt2002,Wang2002,Douarche2005,Garnier2005,Blickle2006,Saira2012,Kung2012} -- systems naturally amendable to performing Landauer-efficient computation. However, a large discrepancy exists between the speed and complexity of the thermodynamically-optimal systems, on one hand, and application-relevant but inefficient traditional processors, on the other. As a consequence, the experimental challenges of operating a Landauer-efficient processor so that its logical functionality and thermodynamic performance are measurable typically preclude complexity beyond one-bit logic. Here, we perform a now-classic Landauer bit erasure experiment on a new hardware platform that promises to obviate many such limitations -- superconducting flux logic~\cite{Chiorescu2003}. It is interesting to note that  recent implementations of heat engines based on weakly anharmonic superconducting resonators~\cite{Cottet2017,Masuyama2018} rely on altogether different operation principles compared to our device which exhibits a strong nonlinearity due to flux quantization.

Exploiting the intrinsic advantages of superconducting flux circuits, our device not only allows for a faithful implementation of the idealized picture put forth by Landauer, but provides a number of practical and theoretical advantages. The magnetic fluxes threading the superconducing loops, though describing macroscopic phenomena, are true microscopic coordinates in the sense that other electronic degrees of freedom are frozen through condensation to a quantum-mechanical ground state. Static controls cause no dissipation on the device, as the magnetic fields are sourced with superconducting leads. The intrinsic clock speed of the system, the plasma frequency, is high ($\omega_p/2\pi \sim 10^{10}$~Hz). Industrial-scale fabrication \emph{en masse} and coupling of a large number of flux logic cells is possible~\cite{King2018,Harris2018}. Owing to these features, high-performance processors implementing complex logical functions have been realized with superconducting architectures~\cite{Burroughs2011,Kirichenko2011,Boixo2014}. For studying the fundamental physics of computing, it is interesting to note that dynamics dominated either by classical or quantum effects can be accessed within this class of devices by a simple change of component values, external bias conditions, or temperature~\cite{Valenzuela2006}. Finally, it is straightforward to engineer the dissipation acting on the remaining dynamical coordinates. Intrinsic dissipation in superconducting circuits has been found to be very low at frequencies up to $\SI{10}{GHz}$~\cite{Turneaure1968,Megrant2012}. Conversely, enhancing dissipation locally is straightforward through the inclusion of resistive normal metal shunts, or coupling to external microwave ports.

In this work, we study information erasure in a gradiometric flux logic cell [Fig.~1(a)]. The information-bearing sub-circuit [Fig.~1(b)] is described by the standard 2D flux qubit Hamiltonian~\cite{Han1989,Han1992,Harris2010}
\begin{gather}
H = \frac{Q^2}{2C} + \frac{Q_{dc}^2}{C/2} + U_0 f(\varphi, \varphi_{dc})\label{eq:H}\\
f(\varphi, \varphi_{dc}) = 
\frac{1}{2} (\varphi - \varphi_x)^2 
+ \frac{\gamma}{2} (\varphi_{dc} - \varphi_{x,dc})^2\nonumber\\
+ \beta_L \cos \frac{\varphi_{dc}}{2} \cos \varphi
+ \delta\beta \sin \frac{\varphi_{dc}}{2} \sin \varphi\label{eq:f}.
\end{gather}
The dynamical coordinates expressed in terms of the junction phases $\delta_1$, $\delta_2$ are $\varphi = \left(\delta_1 + \delta_2\right)/2 - \pi$ and $\varphi_{dc} = \delta_2 - \delta_1$. The bias terms are $\varphi_x = 2\pi \Phi_{ext}/\Phi_0 - \pi$ and $\varphi_{x,dc} = 2\pi \Phi_{ext,dc}/\Phi_0$. $Q$ and $Q_{dc}$ are the common and differential-mode charges on the junction capacitors and conjugate to $\varphi$ and $\varphi_x$, respectively. The potential parametrization is related to the circuit component values as follows: $U_0 = \Phi_0^2 / (4\pi L)$, $\gamma = L/2l$, $\beta_L = 2\pi L (I_{c1} + I_{c2})/\Phi_0$, and $\delta\beta=2\pi L (I_{c2} - I_{c1})/\Phi_0$.

With a suitable choice of the device parameters and the external bias point, the two-dimensional fluxoid potential has the required characteristics for implementing efficient bit storage and erasure. The theoretical potential calculated with calibrated device parameters [Fig.~1(c)] illustrates one of the basic requirements, namely two metastable minima, and the true two-dimensional nature of the system dynamics. In addition, the system allows for independent control of the tilt and the barrier height through the external control fluxes $\varphi_x$ and $\varphi_{x,dc}$. The barrier control has a large tuning range, and allows the potential to be continuously deformed from two effectively isolated wells to a landscape with a single global minimum. We utilize this for device characterization.

\begin{figure}[t!]
    \includegraphics[width=3.375in]{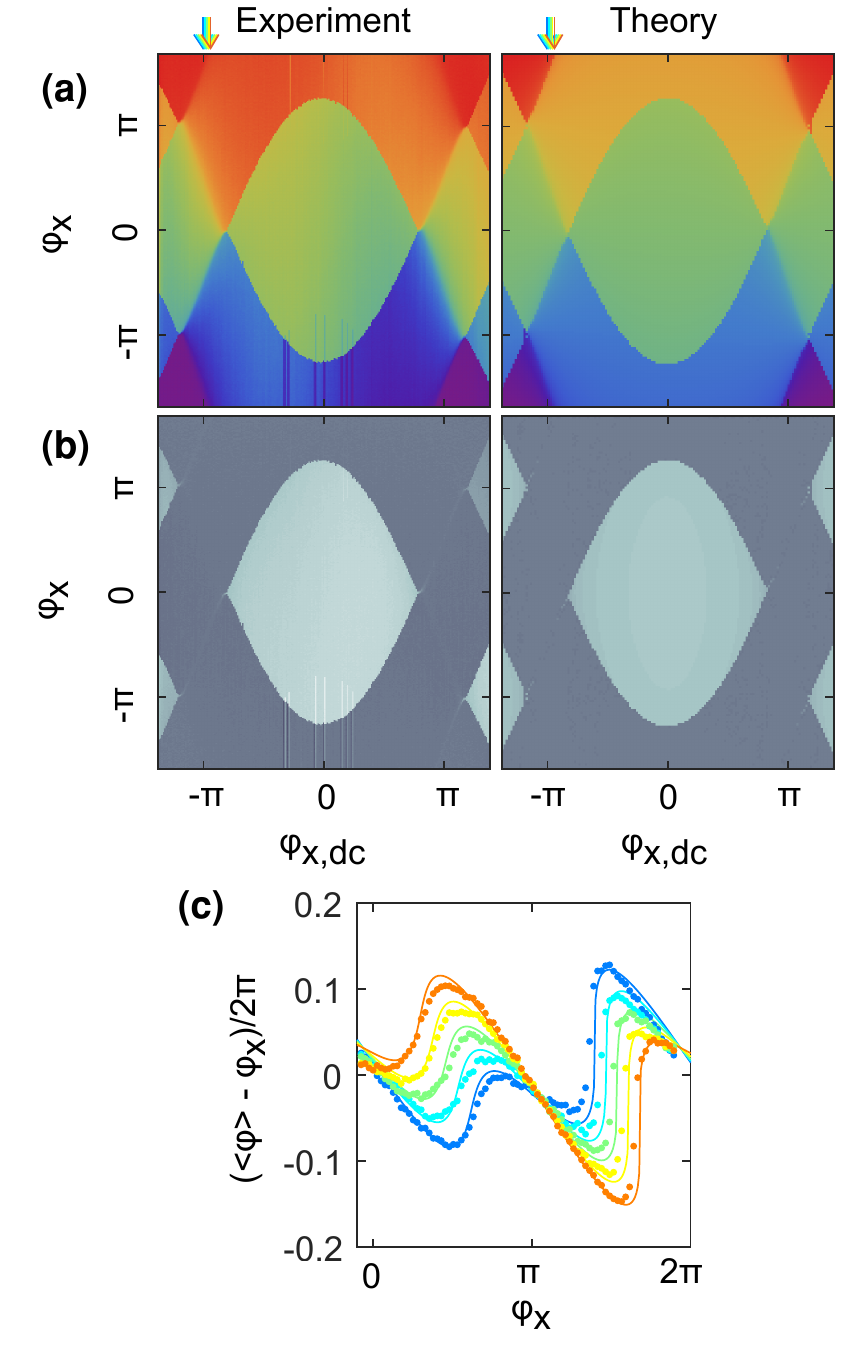}
    \caption{Quasi-static response of the flux logic cell. (a), (b) Two-dimensional scan of tilt and barrier controls ($\Phi_{x}$ and $\Phi_{x,dc}$, respectively). Experimental plots (left) show the unprocessed phase of the local magnetometer readout. Theory plots (right), from Eqs.~(\ref{eq:H}) and (\ref{eq:f}), show the coordinate $\varphi$ at a local minimum of the potential. A bi-directional sweep of tilt was performed for each value of the barrier control. Top panels (a) show the mean response to the two sweep directions. Bottom panels (b) show the response to the positive-direction sweep subtracted from the response to the negative-direction sweep, revealing metastability. (c) Detailed features of the flux response for a few different values of the barrier control [indicated by arrows on top of the (a) panels] in the non-hysteretic regime. Data (markers) and theory (solid lines). The nonlinear flux-to-phase transfer function of the magnetometer has been inverted.}
    \label{fig:2}
\end{figure}
 
Damping, and equilibrium noise associated with it, can be accounted for with a Langevin equation in the classical regime. However, in this work, the characterization experiments as well as the erasure trajectory datasets can be quantitatively explained by a simpler model that only involves the number and location of the critical points of the fluxoid potential. In the subsequent discussion, the labels L, R, and B refer to the two local minima and the saddle point of the $f(\varphi, \varphi_{dc})$ potential landscape, respectively. The right minimum is the one with the larger $\varphi$ coordinate. These points exist and are uniquely defined at all times during the erasure protocols, but not in general. Furthermore, we define $U_i$ as the value of the potential term $U(\varphi, \varphi_{dc})$ at the point $i$, and $U_{ij} = U_j - U_i$. Hence, $U_{LB(RB)}$ gives the barrier height for escape from the metastable minimum L(R), and $U_{LR}$ is the energetic biasing of the double-well system.

The system under study is not overdamped ($Q$-factor evaluated for oscillations in the metastable potential wells is not $< 1$). However, the energy relaxation time $Q / \omega_p = RC$, where $R$ is the effective damping resistance of the logic cell, is much shorter than the timescale over which the external controls are changed, giving the system ample time to equilibrate during the execution of the protocols. This is equivalent to the validity of the Markovian activation-rate description of the inter-well dynamics.

As the first step of the experimental device calibration, we exploit the periodicity of the flux response to set up an affine transformation between the idealized controls $\varphi_x$ and $\varphi_{x,dc}$ and the output voltages of waveform sources that drive the on-chip flux lines through an attenuator network. This transformation is applied implicitly throughout the experiments. For quantitative predictions, a straightforward minima-tracking algorithm reproduces the global behavior of the trapped flux coordinate $\varphi$, including the characteristic $(\Phi_0, 2\Phi_0)$ periodicity in control flux space [Fig.~2(a)], the number of local minima [Fig.~2(b)], and the nonlinear response of the $\varphi(\varphi_x)$ in the single-valued regime [Fig.~2(c)]. We determine the parameter values $\beta_L = 6.2$, $\gamma = 12$, and $\delta \beta = 0.2$ entering Eq.~(\ref{eq:f}) that yield the best agreement with experimental data. We utilize the fact that the mean response is insensitive to the value of $U_0$ and moderate environmental noise when the inter-barrier dynamics are frozen, \ie, $\min\{U_{LB}, U_{RB} \}\gg k_B T$, or there is only one global minimum.

In the case when two minima are separated by a moderate (several times $k_B T$) barrier, the thermally activated Markovian inter-well transition rates are given by~\cite{Li2002,Massarotti2012}
\begin{equation}
    \Gamma_{L,R} = \frac{\Omega}{2\pi} \exp\left(-U_{LB,RB} / E_\mathrm{esc}\right),\label{eq:Rate}
\end{equation}
where $\Gamma_{L,R}$ is the escape rate from well L(R), $\Omega$ is the renormalized plasma frequency, and $E_\mathrm{esc}$ is the escape energy scale. For thermally activated dynamics, $E_\mathrm{esc} = k_B T$. A large body of theoretical results on the nature of information flow and architectural costs in physical computing devices has been derived for systems described by a time-inhomogeneous Markovian model~\cite{Boyd2016,Boyd2018}. Our superconducting device generates faithful realizations of such models in a system for which, moreover, the microscopic dynamics are understood in detail.

\begin{figure}[t!]
    \includegraphics[width=3.375in]{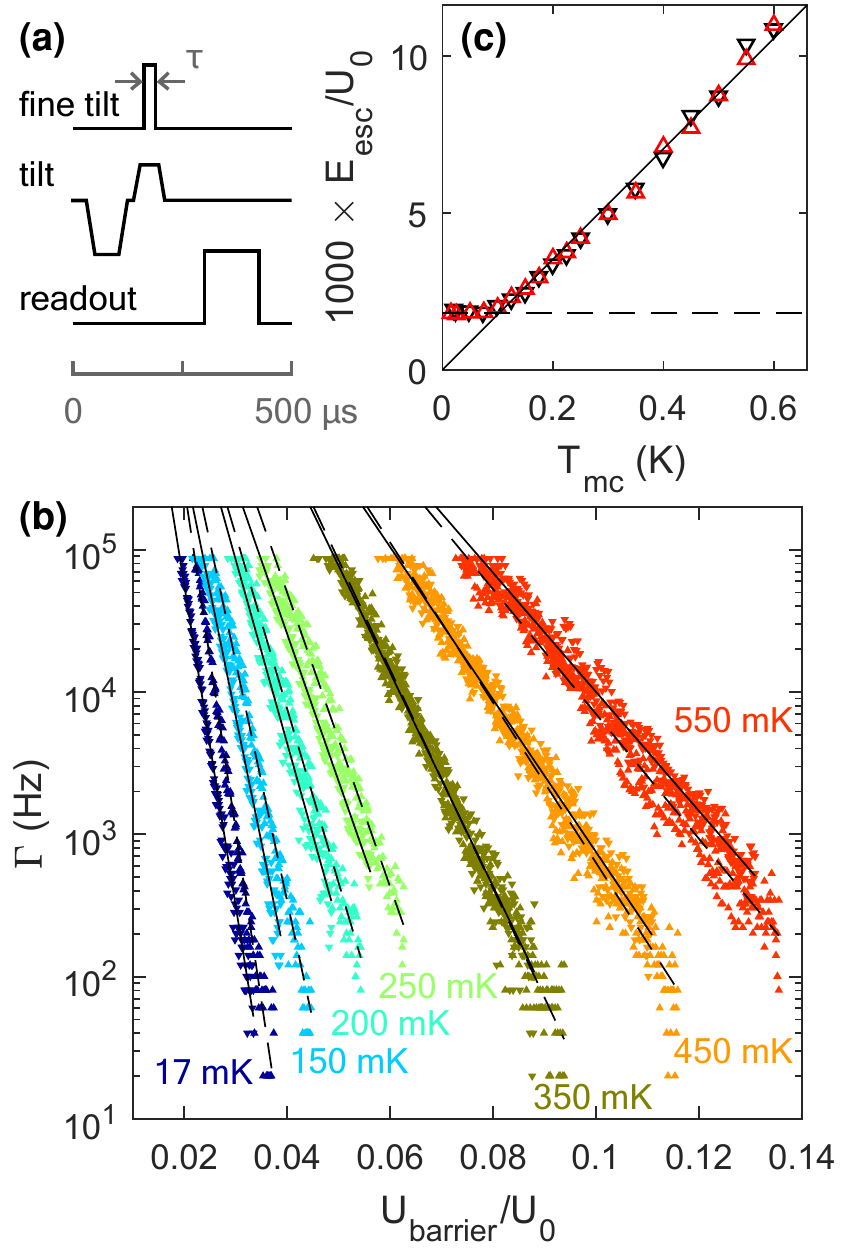}
    \caption{Backaction-free escape dynamics. (a) Pulse sequence used for determining the escape rates. The sequence begins with a deterministic reset. The fine tilt pulse is superimposed with the tilt waveform with a large attenuation. (b) Escape rate as a function of the theoretical barrier height during the fine tilt pulse at different sample temperatures (markers). The experiment was performed at barrier setting $\varphi_{x,dc}/2\pi = -0.3778$ using both positive and negative polarity tilt pulses (upward and downward triangles, respectively). Lines are fits to data (solid and dashed for $+$ and $-$ polarity, respectively). (c) Escape energy, extracted as the inverse negative slope of the escape rate data, as a function of temperature, separately for $+$ and $-$ polarity data (upward and downward triangles, respectively). Solid line is a zero-intercept fit to data at $T > \SI{250}{mK}$, and the dashed horizontal line indicates the low-temperature saturation level.}
    \label{fig:3}
\end{figure}

An important aspect of the technical implementation of the experiment is ensuring that the environmental fluctuations driving the barrier-hopping dynamics correspond to a true thermal bath. In particular, broad-spectrum electromagnetic backaction from the local dc-SQUID magnetometer can cause non-thermal activation above the barrier. To characterize the device dynamics free from magnetometer backaction, we employ a time-domain pulse sequence [Fig.~3(a)] where we make a short excursion of duration $\tau$ to an extreme tilt configuration, thereby probabilistically causing the fluxoid particle to escape to the other minimum. We then return the tilt to the neutral setting ($U_{LR} = 0$), and determine the fluxoid state by a readout pulse. When performed under a sufficiently high barrier, ensuring that the readout pulse does not trigger state transitions at neutral tilt, the escape rate during the maximum tilt can be determined from the observed transition probability $p$ as
\begin{equation}
    \Gamma = -\frac{\log(1 - p)}{\tau}.
\end{equation}

Adherence to the activation rate model [Eq.~(\ref{eq:Rate})] can be verified by determining the escape rate as a function of barrier heights (evaluated from the tilt amplitude and polarity and system parameters) at different temperatures. We have performed the experiment at a constant barrier control $\varphi_{x,dc}/2\pi = -0.3778$ at temperatures up to 600~mK. The data displays the expected exponential dependence of the escape rate on the barrier height~[Fig.~3(b)]. Due to strong asymmetry in the potential, measuring the escape rate in both directions (from Left minimum to Right, and vice versa) serves as an additional check of the validity of the extracted model parameters. Next, we fit the escape energy at each temperature and polarity independently. We find the escape energy to be proportional to the sample temperature above 200 mK~[Fig.~3(c)]. We take this proportionality to be proof of thermally activated dynamics. The common prefactor $U_0$ of the potential can be determined with zero-intercept fit of the dimensionless escape energy in the proportional regime. We obtain $U_0 = k_B\times\SI{56.3}{K}$ (positive polarity) and $U_0 = k_B\times\SI{56.7}{K}$ (negative polarity), corresponding to $L \approx \SI{140}{pH}$. The low temperature saturation corresponds to a temperature $T_{cr} = 103 \pm 2~\si{mK}$ (positive polarity) and $T_{cr} = 105 \pm 2~\si{mK}$ (negative polarity).

A fundamental explanation of the low-temperature saturation is the transition from thermally activated dynamics to macroscopic quantum tunneling (MQT). Within this interpretation, $T_{cr} = \hbar\omega_p / (2\pi k_B)$~\cite{Affleck1981,Grabert1984,Hanggi1985}, and the plasma frequency $\omega_p / 2\pi$ of the system is approximately $\SI{13.7}{GHz}$ at the operation point used for the escape rate experiments. Alternatively, we can estimate $C$ from the total junction area ($\SI{11.8}{{\mu}m^2}$, based on a high-magnification SEM image) and the nominal specific capacitance $\SI{45}{fF/{\mu}m^2}$ of the junction fabrication process. This yields $C = \SI{530}{fF}$ and $\omega_p/2\pi = \SI{18.5}{GHz}$. The 30\% relative discrepancy in estimated $\omega_p$ will not affect the conclusions we make. We use the lower value given by the MQT experiment for the remainder of our analysis here. Macroscopic resonant tunneling~\cite{Rouse1995} peaks were not resolvable, presumably due to small level separation in terms of tilt flux ($4.7 \times 10^{-4} \Phi_0$). All further experiments are performed at a temperature of 500~mK, firmly in the thermal activation regime.

\begin{figure}[t!]
    \includegraphics[width=3.375in]{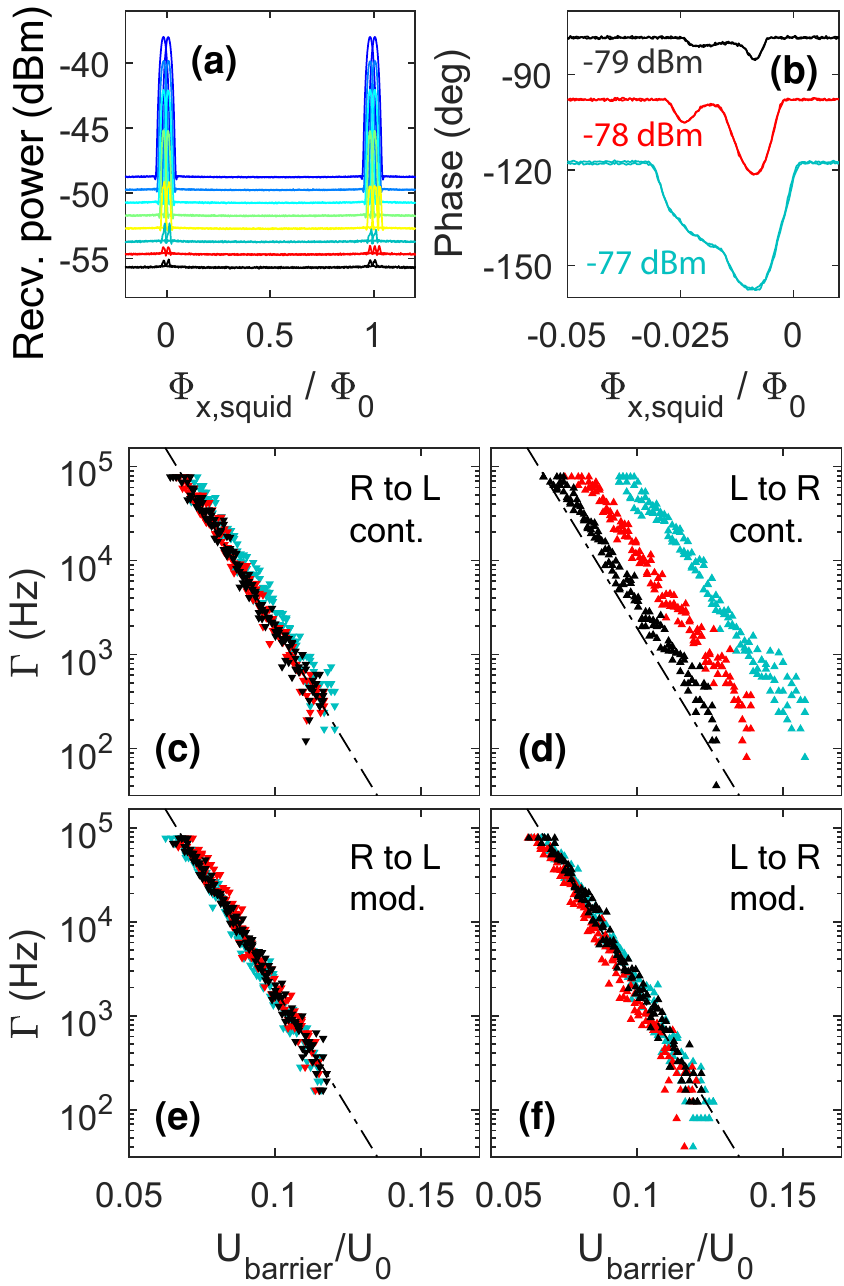}
    \caption{Magnetometer readout with different probe powers and backaction on system dynamics at $T = \SI{500}{mK}$. (a) Flux modulation curves for $-$79~dBm to $-$72 dBm incident power, in 1 dBm steps. Pickup from magnetometer bias to tilt and barrier fluxes has not been compensated. (b) Phase response for the three lowest powers. Data has been offset by 20$^\circ$ per dBm for clarity. (c)-(f) Escape rates at $\varphi_{x,dc}/2\pi = -0.3778$ under continuous readout [(c), (d)] versus backaction-free pulsed readout [(e), (f)] for the three lowest powers. Operation point has been chosen so that the logical state R is at the peak of the modulation curve. Consequently, the magnetometer is (close to) a zero voltage state in the logical state L. Dash-dotted line is the same in all panels and serves as a guide for the eye.}
    \label{fig:4}
\end{figure}

To extract work statistics for bit erasure at the single trajectory level, the system state must be continuously tracked throughout the protocol. Therefore, we characterize the logical state-dependent backaction of the local magnetometer when read out with a continuous low-power sinusoidal $(f = \SI{10}{MHz})$ waveform. The probing frequency was chosen so that it is in the passband of an ac-coupled cryogenic SiGe preamplifier. The preamplifier output was demodulated with an RF lockin-amplifier into two zero-IF quadrature channels that were subsequently digitized. The magnetometer SQUID was not part of a resonant circuit. Hence, the observed flux modulation characteristics [Fig.~4(a), (b)] as well the nature of its backaction on the flux logic cell resemble those of a current-biased dc-SQUID. From similar data, we extract magnetometer flux shift due to the logical transition at $\varphi_{x,dc} = 0$ to be $0.024 \Phi_0$, and the mutual inductance between the magnetometer and the flux logic cell $M = 0.0254 L \approx \SI{3.6}{pH}$. From a separate low-temperature dc four-write characterization of the readout SQUID, we determine its shunt resistance $R_\mathrm{shunt} = \SI{2.1}{\Omega}$. We estimate the $Q$-factor due to shunt-induced damping of the flux dynamics as~\cite{Neely2008}
\begin{equation}
Q_\mathrm{shunt} \approx \frac{R_\mathrm{shunt} L}{\omega_p M^2} = 260,
\end{equation}
where we have used the approximation $\omega_p \approx 1/\sqrt{L C}$, thus neglecting the contribution from the Josephson inductance. Noting that the chip contains two nominally identical readout circuits placed symmetrically with respect to the logic cell, we obtain an upper bound on the inter-well relaxation timescale $Q / \omega_p \leq Q_\mathrm{shunt}/2\omega_p = \SI{9.5}{ns}$, where the total $Q$ includes all damping mechanisms. This sets the fundamental limit on what constitutes an adiabatic evolution in the system. However, the reset protocols studied in this manuscript are many orders of magnitude slower, ensuring that the timing of the logical state transitions can be accurately determined from the finite-bandwidth magnetometer output. 

The potentially harmful nonequilibrium backaction from a dc-SQUID appears in the form of wideband microwave radiation with a complex spectrum peaked at $\omega_J = 2eV/\hbar$ and harmonics, where $V$ is the dc voltage developed over the junction. For low-amplitude probing currents, it is possible to choose the magnetometer flux bias in such a way that the SQUID is in a finite-voltage state for logical state L and in the zero-voltage state for logical state R. This configuration would be expected to result in logical-state-dependent backaction. We quantify the backaction by repeating the earlier escape-rate experiment with both continuous and pulse-modulated readout at $T = \SI{500}{mK}$. The data [Fig.~3(c)-(f)] is in agreement with the model of readout backaction laid out above. We find only the L-to-R escape under continuous readout [Fig.~3(d)] to be affected. The escape rate appears to be enhanced by a constant power-dependent factor, but the effective temperature, quantified by the slope of the $\Gamma$ vs. $U_{barrier}$ characteristic, is not affected. Guided by this characterization, we choose an incident readout power of $-79$~dBm (to 50~$\Omega$ load) for the continuous monitoring of stochastic bit erasure trajectories. At this power level, the rate enhancement is equivalent to a sub-$k_B T$ change in the energetics of the system, while the signal-to-noise remains sufficient for fast discrimination of the logic state~[Fig~5(c)].

\begin{figure*}
    \includegraphics[width=7in]{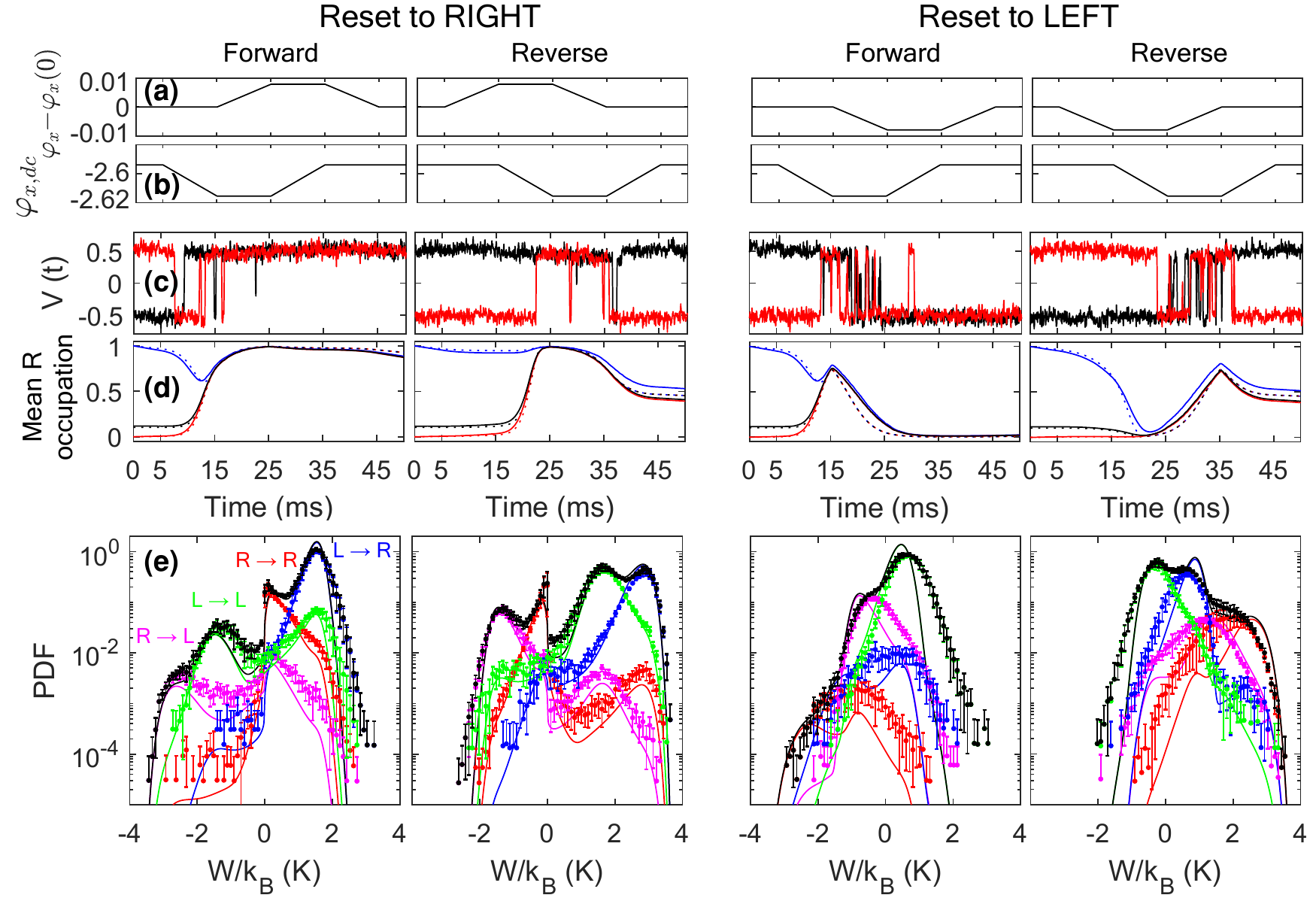}
    \caption{Bit erasure at $T = \SI{500}{mK}$. (a), (b) Time-domain waveforms applied to tilt and barrier controls. The reverse protocol is obtained by reversing both control channels in time. The reset protocol targeting state R has the polarity of the tilt waveform inverted. (c) Two randomly chosen magnetometer traces. (d) Average occupation of the state L throughout the protocol when the initial state is L (blue), R (red), or an equilibrium mixture of the idling potential (black). Experimental average from $10^5$ trajectories (solid) and theory (dashed). (e) Distribution of work $W$ from $10^5$ experimental trajectories (markers) and from Markovian theory (lines). Full distribution (black) and conditional distributions based on initial and final states (magenta, green, red and blue for trajectories of R $\rightarrow$ L, L $\rightarrow$ L, R $\rightarrow$ R and L $\rightarrow$ R type, respectively) are shown.}
    \label{fig:5}
\end{figure*}

\begin{figure}[t!]
    \includegraphics[width=3.375in]{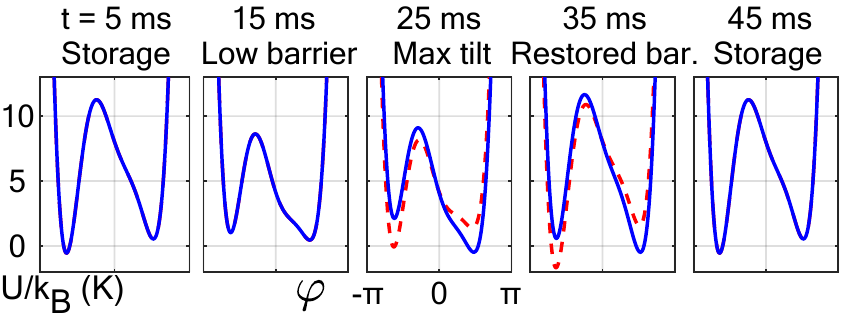}
    \caption{One-dimensional projection of the potential at key stages of the Reset-to-R (blue) and Reset-to-L (red, dashed) protocols. The plotted quantity is the potential as a function of the $\varphi$ coordinate evaluated along a curve in $(\varphi, \varphi_{dc})$ space that passes through the left minima, the saddle point, and the right minima (in this order), and is parallel to the local principal axis of curvature at these points. A constant offset of $\SI{71.4}{K}$ has been subtracted for clarity.}
    \label{fig:1d}
\end{figure}

To study the work statistics of bit erasure, we implement the reset protocol used in Ref.~\cite{Jun2014}. The protocol starts from and ends in a storage state. Logical state reset is realized by piecewise linear controls applied to the tilt and barrier channels. Efficient and fast reset is achieved by changing the controls in a particular sequence: Drop~barrier--Tilt--Raise~barrier--Untilt [Figs.~5(a),(b) show the waveforms; Fig.~\ref{fig:1d} shows the induced potentials]. With the control waveform shapes fixed, one still has a choice of their duration, flux offset, amplitude, polarity, and directionality.

We derive three transformed versions of the basic protocol that implements Reset-to-R functionality: Reset-to-L, obtained by inverting the polarity of the tilt waveform while maintaining the same offset: $\varphi_x^L(t) = \varphi_x(0) - [\varphi_x^R(t) - \varphi_x(0)]$. And, for both polarities, the reversed protocol is obtained by time-reversing both the tilt and barrier waveforms. Importantly, due to the finite $\delta \beta$ term in the Hamiltonian [Eq.~(\ref{eq:H})], reversing the sign of the tilt control $\varphi_x$ and the longitudinal coordinate $\varphi$ does not result in an equivalent potential landscape. Hence, the Reset-to-L and Reset-to-R protocols give rise to a different distribution of microscopic trajectories. Snapshots of the potential at key stages of the reset protocols are shown in Fig.~\ref{fig:1d}. 

In the infinite time limit, an application of this protocol with appropriate scaling of the controls results in a Landauer-efficient reset with Gaussian work statistics. For a finite-duration protocol with ideal control of the energetics, the resulting work histogram is bimodal and displays characteristic features that can be traced back to different sub-stages of the protocol~\cite{Wimsatt2019}. In our experiment, the asymmetry of the Hamiltonian leads to a nontrivial functional dependence of the energetics (in particular, $U_{LR}$) on the external controls $(\varphi_x, \varphi_{x,dc})$, giving rise to complex multimodal work distributions [Fig.~\ref{fig:5}(e)].

The important timescales of the experiment were chosen to satisfy
\begin{equation}
\tau_{\mathrm{readout}} \ll \Gamma_{\mathrm{tilt}}^{-1} \ll \tau_{\mathrm{total}},
\end{equation}
where $\tau_{\mathrm{readout}} \approx \SI{0.1}{ms}$ is the time needed to detect a logical transition, $\Gamma_{\mathrm{tilt}}$ is the typical transition rate of the system during the Tilt phase ($25\ldots\SI{35}{ms}$), and $\tau_{\mathrm{total}}$ is the total duration of the protocol. The first condition ensures that the magnetometer can track the system dynamics, and the second ensures that the system can sample both wells during the protocol execution. For the protocols studied here, $\Gamma_{\mathrm{tilt}} \approx \SI{1}{kHz}$. Finally, we choose $\tau_{\mathrm{total}} = \SI{50}{ms}$. This allows us to collect $N = 10^5$ trajectories for all four protocol transformations with a total measurement time of $\SI{12}{h}$ with 50~\% duty cycle. For the duration of the acquisition, the system is run with open-loop control.

We monitor the magnetometer output continuously during the execution of each reset protocol. Two randomly chosen traces are shown in Fig.~5(c). The polarity of the magnetometer response is such that the Left logical state corresponds to a positive output voltage. We classify the instantaneous logical state according to the sign of the magnetometer signal. Evaluating the mean occupation of either logical state (here, we choose R) as a function of time results in, over the $10^5$ experimental trajectories, a smoothly varying curve [Fig.~5(d)].

\begin{figure}[t]
    \includegraphics[width=3.375in]{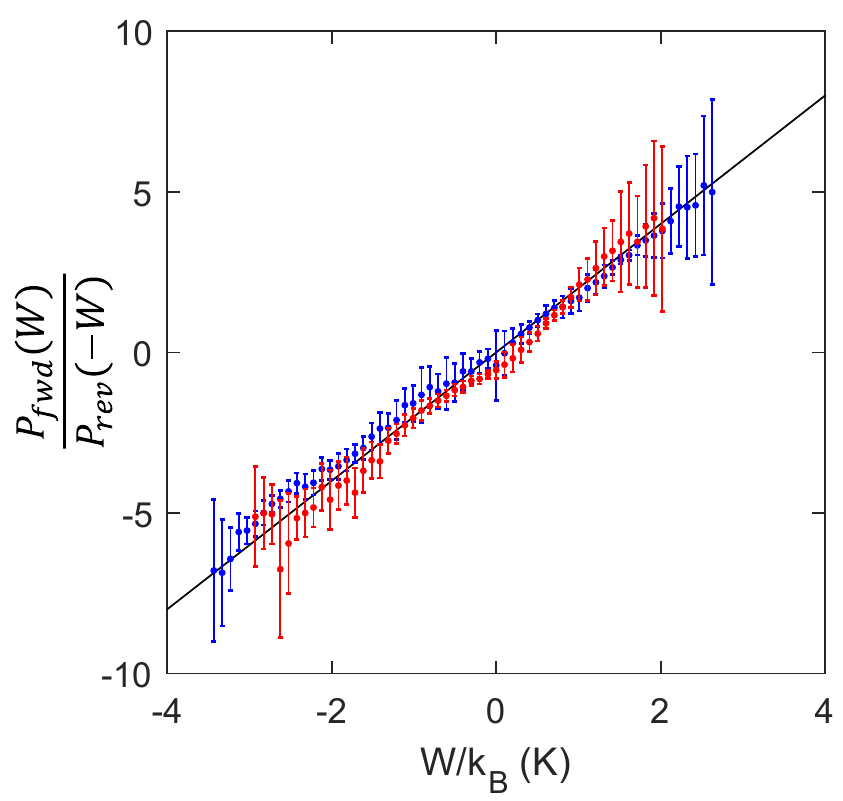}
    \caption{Crooks relation ratio of the probability densities of per-trajectory work in forward and reverse directions (see text for definitions). Experimental ratio for bit reset targeting the R and L states (blue and red  markers, respectively), and the Fluctuation Theory result $\exp\left(W/k_B T\right)$ with $T = \SI{500}{mK}$ (solid line).}
    \label{fig:ratio}
\end{figure}

Evidently, logical state transitions correspond to the zero-crossings in the magnetometer signal. We will use this symbolic representation of the system state to derive a per-trajectory work estimate. Consider a trajectory that starts in state $s_0 \in \{L, R\}$ and involves $n$ state transitions such that the $i$th transition occurs at time $\tau_i$ and takes the system to state $s_i \in \{L, R\}$. Defining $\tau_{n+1} = \tau_\mathrm{total}$, we write the per-trajectory work as
\begin{eqnarray}
    W & = & \sum_{i=1}^n \left[U_{s_i}(\tau_i) - U_{s_i}(\tau_{i-1})\right]\\
      & = & \left[U_{s_n}(0) - U_{s_0}(0)\right] + \sum_{i=1}^n \left[ U_{s_{i}}(\tau_i) - U_{s_{i-1}}(\tau_{i})\right].\label{eq:W}
\end{eqnarray}
The second equality makes use of the fact that the potentials at $t = 0$ and $t = \tau_\mathrm{total}$ are identical. The latter form illustrates that $W$ can be expressed as a sum of $U_{LR}(\tau_i)$ terms with alternating signs.

Even though our experimental flux traces consist of discrete fluxoid state transitions, the underlying dynamical coordinates are continuous. Evaluation of Eq.~(\ref{eq:W}) gives an accurate estimate of the true microscopic work, provided: (i) two metastable minima exist throughout the protocol; (ii) the system has time to equilibrate between logical transitions; (iii) the changes in control parameters are slow compared to the internal equilibration time; and (iv) the potential landscape is only weakly perturbed. The details of this argument in the context of superconducting flux logic are laid out in Ref.~\onlinecite{Wimsatt2019}. An equivalent approach is commonly used in studies of nonequilibrium thermodynamics in single-electron devices~\cite{Kung2012,Koski2014}. Numerical Langevin simulations of a double-well system satisfying the above conditions confirm that the work distribution evaluated with the discretized formula in Eq.~(\ref{eq:W}) agrees with that obtained for the microscopic work evaluated with continuous coordinates.

For the parameters of this experiment, the experimental initial tilt offset and the asymmetry of the potential give rise to a nonzero $U_{LR}(0)/k_B = -\SI{1.01}{K}$. We determine this initial energy offset based on the equilibration of the left and right state populations during the first 1/10th of the protocol, for which the system is in the storage state. Conversely, choosing a weighting for left and right initial conditions based on the Boltzmann factor corresponding to this energy offset results in a steady occupation in the initial idling period [Fig.~5(d), black line]. We use the same Boltzmann-factor weighing when aggregating the work histograms. With this weighting, Fluctuation Theorems are satisfied by the quantity $W$ defined above. Note that the first term of Eq.~(\ref{eq:W}) vanishes if the potential in the initial storage state is degenerate, \ie, if $U_{LR}(0) = 0$. 


The per-trajectory work estimate is based on the calibrated potential, but does not require a model of the system dynamics. Given that we, in addition, calibrated the two-state activation rate model, we can use that knowledge to predict the system's time-domain response to the erasure protocols. The renormalization of the rates due to damping and local curvature of the potential~\cite{BenJacob1983,Han1989,Li2002} is a much smaller effect than the variance due to uncertainty in the model parameters. Consequently, we substitute $\Omega = \omega_p$ as the prefactor in Eq.~(\ref{eq:Rate}). Such predictions are included in the mean occupation plots of Fig.~5(d) as dashed lines, and the work histograms of Fig.~5(e) as solid lines. This simple dynamical model reproduces with good accuracy the mean occupations throughout the protocols as well as the locations and relative weights of the peaks in the multimodal work histograms. One can easily discern, however, that the agreement is worse for the Reset-to-L family of protocols. The reason for the disagreement is not clear at the moment, but similar features can be observed in other datasets acquired from the same device. 

To quantify the effect of slow flux offset drifts, we process the data in 10 chunks of $10^4$ consecutive trajectories, and plot the mean and $2\sigma$ confidence intervals for each bin of the work histograms [Fig.~5(e)]. Left and right initial conditions have been weighted according to the Boltzmann factor defined above. We include the per-bin uncertainties in the evaluation of the Crooks-relation~\cite{Crooks1999} ratios $\rho = P_{fwd} (W) / P_{rev}(-W)$ using standard error propagation formulas [Fig.~7]. When evaluated in this manner, the confidence intervals also include the statistical uncertainty due to finite sampling, dominating the uncertainty for low-count bins. The fact that the Crooks-relation ratios fall on the expected line $\log(\rho) = k_B T$ within the error bars is another indication that our model of the microscopic energetics of the flux logic system is correct.

In conclusion, we presented a trajectory-level analysis of the thermodynamics of information erasure in a superconducting flux logic device, where a double-well potential arises naturally through a combination of the Josephson effect and flux quantization. We calibrated a microscopic model of the device energetics and evaluated detailed work histograms for bit erasure protocols in a parameter regime where metastable two-state approximation is valid throughout the protocol. We also demonstrated that a simple dynamical model, based on the calibrated potential and barrier activation, explains all experimental observations in detail. This sets the stage for designing and diagnosing thermodynamically efficient computing based on superconducting devices.

In this initial study, the execution speed of the bit reset was constrained by the limitations of the dc-SQUID readout scheme. Future experiments employing either dispersive readout~\cite{Quintana2017} with a wideband quantum-limited preamplifier~\cite{HoEom2012} or thermal detectors~\cite{Pekola2013} will enable GHz-scale clock rates while still maintaining fraction-of-$k_B T$ excess dissipation, and a similar resolution for the extracted thermodynamical quantities.

\emph{Acknowledgments:} We thank A. Boyd and C. Jarzynski for helpful discussions. As an External Faculty member, JPC thanks the Santa Fe Institute and JPC, OPS, MHM, RK, WF, GW, and MLR thank the Telluride Science Research Center for their hospitality during visits. This material is based upon work supported by, or in part by, the U. S. Army Research Laboratory and the U. S. Army Research Office under contracts W911NF-13-1-0390 and W911NF-18-1-0028.




\bibliography{FluxLogic}

\end{document}